\documentstyle[12pt,epsfig]{article}
\newcommand{\be}{\begin{equation}}
\newcommand{\ee}{\end{equation}}

 1
\font\elevenrm=cmr10 scaled\magstep 1
 1

\def\reff{\hang\noindent}

\begin{document}
\vspace*{1.8cm}
  \centerline{\bf OPENING THE ULTRA HIGH ENERGY COSMIC RAY WINDOW} 
  \centerline{\bf FROM THE TOP}
\vspace{1cm}
  \centerline{PASQUALE BLASI}
\vspace{1.4cm}
  \centerline{NASA/Fermilab Astrophysics Center}
  \centerline{\elevenrm Fermi National Accelerator Laboratory, Box 500,
Batavia, IL 60510-0500}
\vspace{3cm}
\begin{abstract}
While several arguments can be proposed against the existence of particles 
with energy in excess of $(3-5)\times 10^{19}$ eV in the cosmic ray spectrum,
these particles are actually observed and their origin seeks for an 
explanation. After a description of the problems encountered in explaining
these ultra-high energy cosmic rays (UHECRs) in the context of astrophysical
sources, we will review the so-called {\it Top-Down} 
(TD) Models, in which UHECRs are the result of the decay of very massive 
unstable particles, possibly created in the Early Universe. 
Particular emphasis will be given to the signatures of the TD models,
likely to be accessible to upcoming experiments like Auger.
\end{abstract}
\vspace{2.0cm}

\section{Introduction}
Cosmic ray particles with energy in excess of $\sim 10^{20}$ eV have been 
detected during the last thirty years by several independent experiments,
such as AGASA (Takeda et al. 1998; Takeda et al. 1999; Hayashida et al. 1994),
Fly's Eye (Bird et al. 1993, 1994, 1995), Haverah Park (Lawrence, Reid and 
Watson 1991), Yakutsk (Efimov 1991), Volcano Ranch (Linsley 1963) and more
recently by the High Resolution Fly's Eye experiment (Kieda et al. 1999).  
These events represent now more than ever a big challenge for 
our understanding of particle physics and astrophysics. 

While hystorically the first reactions to the detection of these particles
were related to the already difficult problem of accelerating particles to
the highest observed energies, it became soon clear that the existence of 
cosmic rays having energy larger than $\sim 4\times 10^{19}$ eV [the 
so-called ultra-high energy cosmic rays (UHECR)] was more than that, and
indeed represented a much more serious challenge to known Physics. Soon after 
the discovery of the cosmic microwave background radiation (CMBR), 
Zatsepin and Kuzmin (1966) and independently Greisen (1966) recognized that
the propagation of a proton in the CMBR bath had to be limited to short 
distances due to photopion production. If the sources of UHECRs are 
distributed homogeneously in the sky, this immediately implies that the 
flux of UHECRs above $\sim 4\times 10^{19}$ eV should be strongly suppressed.
This is the so-called Greisen-Zatsepin-Kuzmin (GZK) cutoff. Similar arguments
apply to nuclei. 

This puzzling situation inspired on one side a proliferation of models of the 
generation of particles with sufficiently high energy, and on the other
side it fueled interest in the study of the propagation of UHECRs 
[see for instance Lee, Olinto and Sigl 1995; Lemoine, Sigl, Olinto and 
Schramm 1997 and Bhattacharjee and Sigl 2000, and references therein] for 
different compositions and for realistic models for the 
distributions of the sources and for the intergalactic magnetic field, which 
still remains only contrained by upper limits, generally based on
measurements of the Faraday rotation of light coming from distant 
quasars (Kronberg 1994; Blasi, Burles and  Olinto 1998). These limits are 
at the level of $\sim 10^{-9}$ Gauss, although larger values are allowed
in large scale structures. The angular deflection of ultra high 
energy protons in such fields would be comparable with or smaller than 
the angular resolution of current experiments, so that in principle it 
should be possible to do astronomy using UHECRs as probes. Several efforts 
have been put into the search for candidate nearby sources in
the direction of arrival of UHECRs, but with no result (see for instance
Elbert and Sommers 1995).

Recent analysis of the distribution of arrival directions of UHECRs
(Takeda et al. 1999; Uchihori et al. 2000), in
search for a possible large scale anisotropy also gave negative results: 
with the present statistics, the observed distribution appears to be
consistent with isotropy, but indications have been found of small scales
anisotropies, at a few degrees level. If confirmed, this finding will hopefully
provide hints about the sources of UHECRs.

The paper is structured as follows: in section 2 we give a short outline
of the observational situation; in section 3 we present a critical view
of the GZK cutoff, its meaning and its potential power in limiting wide classes
of models. In section 4 we introduce the astrophysics and particle physics
inspired models; in section 5 we discuss topological defects 
variants of TD models, while in section 6 we discuss the models of relic
quasistable massive particles. We conclude in section 7.

\section{Observations}

The cosmic ray spectrum is measured from fractions of GeV to a (current)
maximum energy of $3\times 10^{20}$ eV. The spectrum above a few GeV and up to 
$\sim 10^{15}$ eV (the knee) is measured to be a power law with slope
$\sim 2.7$, while at higher energies and up to $\sim 10^{19}$ eV 
(the ankle) the spectrum has a different slope, of $\sim 3.1$. At energy 
larger than $10^{19}$ eV a flattening seems to be present. 

There are currently 92 events above $4\times 10^{19}$ eV, 47 of which 
have been detected by the AGASA experiment.

The information available on the composition of cosmic rays at the highest 
energies is quite poor. A study of the shower development was possible only 
for the Fly's Eye event (Bird et al. 1995) and disfavors a photon primary 
(Halzen et al. 1995). A reliable analysis of the composition is however 
possible only on statistical basis, because of the large fluctuations 
between one shower and the other at fixed type of primary particle.

The Fly's Eye collaboration reports of a predominantly heavy composition
at $3\times 10^{17}$ eV, with a smooth transition to light composition
at $\sim 10^{19}$ eV (see talk by Alessandro, these proceedings). 
This trend was not confirmed by AGASA (Hayashida et al.
1994; Yoshida and Dai 1998).

In (Takeda et al. 1999) the directions of arrival of the AGASA events above 
$4\times 10^{19}$ eV were studied in detail: no appreciable departure 
from isotropy was found, with the exception of a few small scale 
anisotropies in the form of doublets and triplets of events within 
an angular scale comparable with the angular resolution of the experiments 
($\sim 2.5^o$ for AGASA). This analysis was repeated in (Uchihori et al. 2000)
for the whole sample of events above $4\times 10^{19}$ eV, and a total of 12 
doublets and 3 triplets were found within $\sim 3^o$ angular scales. 
The attempt to associate these multiplets with different types of astrophysical
sources possibly clustered in the local supercluster did not give positive
result (see Stanev, these proceedings). 

\section{The GZK cutoff: what is it telling us?}

Although the existence of UHECRs is experimentally well established, it 
represents a big challenge from the theoretical point of view, because of 
a combination of puzzles related to the production and to the propagation 
of these particles. We will summarize the different parts of this puzzle 
in the following. 

\begin{figure}[thb]
 \begin{center}
  \mbox{\epsfig{file=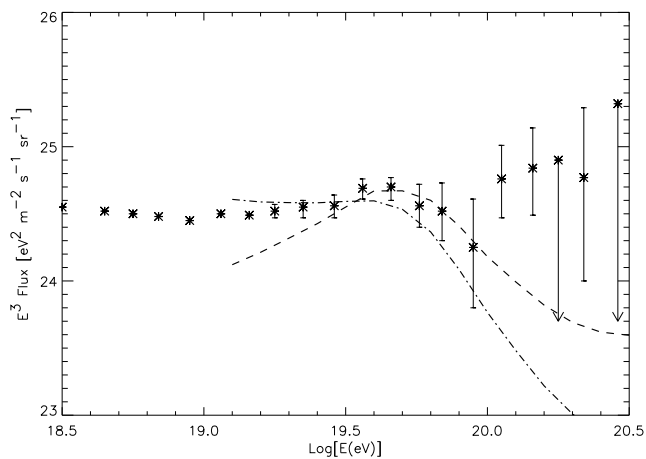,width=9.cm}}
  \caption{\em {Spectrum of UHECRs detected by AGASA. The lines are 
theoretical predictions for homogeneously distributed sources and injection
spectra $E^{-2}$ (dashed line) and $E^{-3}$ (dash-dotted line). 
}}
 \end{center}
\end{figure}

The first and most important part of the mystery is that the Universe should
be dark at energies in excess of a few $10^{19}$ eV due to photopion 
production of UHECRs on the photons of the CMBR. This 
interaction has a typical pathlength $l_{int}< 50$ Mpc (corresponding
to a travel time of $< 10^8$ years) at ultra high 
energies (see Stanev, these proceedings). If the sources of UHECRs are 
distributed nearly homogeneously in the universe,
the photopion production results in an observed spectrum which has a 
pronounced cutoff that starts at $\sim 3\times 10^{19}$ eV. This is the 
so-called GZK cutoff (Greisen 1966; Zatsepin and Kuzmin 1966). 
It is worth spending a few more words on the meaning of the cutoff, since
it is so crucial in defining the problem of UHECRs. Particles generated
at distances closer than $\sim l_{int}$ can reach the detector and be 
UHECR events. The pathlength $l_{int}$ becomes gradually larger at lower
energies, so that particles with these energies can come from correspondingly
larger distances. It is clear that the problem of UHECRs
exists because, for a homogeneous distribution of sources there are 
not enough nearby sources to provide the observed fluxes. In Fig. 1 we show
the results of AGASA observations at energies $>10^{18.5}$ eV and the 
theoretical
prediction for a homogeneous distribution of sources and an injection spectrum
$E^{-2}$ (dashed line) and $E^{-3}$ (dash-dotted line). 
Two comments are in order on this figure: 1) the predicted 
spectra may present a recovery above some energy, due to the flux contributed 
by the nearby sources; 2) the energy position of the cutoff increases for 
a locally overdense distribution of sources.

Only if the universe presents local overdensities by a factor $> 10$
it is possible to reconcile the expected spectra with the observed spectra and
fluxes (Berezinsky et al. 1990), provided suitable sources are found in 
the local universe. These overdensities are not observed (Blanton, Blasi 
and Olinto 2000).

At this point the second part of the mystery enters: upper limits on the 
intergalactic magnetic field based on the Faraday rotation measurements
are at the level of $10^{-9}-10^{-10}$ Gauss (Kronberg 1994; Blasi et al. 
1999), so that typical deflection angles of protons are forced to be 
within a few degrees. 
Searches for sources of UHECRs have been carried out, but no plausible 
candidate was found within $\sim 3^o$ of the direction of arrival of the 
events (Elbert and Sommers 1995). If the candidate
particle is an iron nucleus then the deflection angles are likely to become 
larger and it becomes correspondingly harder to look for sources. However, 
if the deflection angle becomes too large, then the regime of propagation 
becomes closer to diffusive and this forces the distance to the source to 
be even smaller, making the situation more problematic. A possible exception
to this is if the iron nuclei are accelerated locally in the Galaxy and
deflected and isotropized in an extended halo. In this case no GZK cutoff 
is expected.

The problems mentioned above are all related to the propagation of UHECRs.
However there is another problem, that hystorically was the first to be
studied, and is related to the mechanism able to produce such particles. 
We dedicate the next section to this topic.

\section{Astrophysics and Particle Physics Models}

Astrophysical models for the production of UHECRs are generally based on 
an acceleration mechanism to be applied to a class of scenarios of 
astrophysical relevance. We do not discuss here any of the acceleration models,
but we outline the general features and problems they encounter.
An exhaustive discussion of most of the models 
which are currently under investigation can be found in (Olinto 2000) and some
of them are discussed by Stanev (these proceedings). A very broad
classification of these models is that between galactic and extragalactic 
ones. In general, galactic models require heavy composition, in order to 
emulate the isotropic distribution of arrival directions. 

In most of the extragalactic models that have been proposed so far, the 
highest energies are reached only for the extreme values of the parameters 
involved [see (Norman, Melrose and Achterberg 1995)]. 
Strong constraints on astrophysical models 
come from the spectrum they generate (after accounting for propagation).
In fact, as long as the acceleration models are associated with sources with 
a homogeneous distribution in the sky, the problem of UHECRs 
remains, independently on the maximum energy. The case of a single source 
accidentally present in the nearby universe (for instance in the local 
supercluster) was investigated by Berezinsky, Grigorieva and Dogiel 1990; 
Blasi and Olinto 1999; Sigl, Lemoine and Biermann 1999). 
Even with relatively strong fields of $10^{-8}-10^{-7}$ Gauss, the anisotropy 
is too large compared with observations (see however (Ahn et al. 1999) for 
an alternative model involving M87).

The difficulty in acceleration to the highest energies (Norman, Melrose and
Achterberg 1995), the presence of
the GZK cutoff in most of the cases and the lack of counterparts in the 
arrival directions fueled the interest in a new class of models that 
could avoid these problems. These are {\it particle physics inspired models},
in which UHECRs are generated as a result of the decay of very massive
particles (from here the name of Top-Down models). The problem of reaching 
the maximum energies is, in these models, solved {\it by construction}. 

The spectra of the particles resulting from the decay are determined 
in principle by QCD, the channel of reactions being: $X\to q\bar q$, where
$q$ are quarks that hadronize, generating mainly pions and a small 
fraction of protons and neutrons. The spectra are generally very flat 
(roughly $E^{-3/2}$, although the realistic calculations do not give power law
spectra), which represents one of the peculiar features 
of TD models. At the production, most of the ultra-high energy particles
are gamma rays, but propagation effects can change the ratio of gamma
rays to protons. The gamma rays generated at distances larger than the 
absorption length produce a cascade at low energies (MeV-GeV) which 
represents a powerful tool to contrain TD models.

There are basically two ways of generating the very massive particles
and make them decay at the present time: 1) trapping them inside topological 
defects; 2) making them quasi-stable (lifetime larger than the present age 
of the universe) in the early universe. 
We discuss these two possibilities separately in the next two sections.

\section{Topological Defects}

Topological defects are naturally formed at phase transitions and
their existence has been proven by direct observations in several 
experiments on liquid crystals and ferromagnetic materials. 
Similar symmetry breakings at particle physics level are responsible for the
formation of cosmic topological defects [for a review see (Vilenkin and
Shellard 1994)]. 

The fact that topological defects can generate UHECRs was first proposed in
the pioneering work of Hill, Schramm and Walker (1987). The general idea 
is that the stability of the defect can be locally broken by different types of
proceses (see below): this results in the false vacuum, trapped within the 
defect, to fall into the real vacuum (outside universe), so that the gauge
bosons of the field trapped in the defect acquire a mass $m_X$. 
At this point, the very massive and unstable particles rapidly decay 
producing high energy particles. 

Several topological defects have been studied in the literature: ordinary
strings (Bhattacharjee and Rana 1990), superconducting strings (Hill, Schramm
and Walker 1987), bound states of magnetic monopoles (Hill 1983; Bhattacharjee
and Sigl 1995), networks of monopoles and strings (Berezinsky, Martin and
Vilenkin 1997), necklaces (Berezinsky and Vilenkin 1997) and vortons (Masperi 
and Silva 1998).

Only strings, necklaces and monopolonia will be considered here, while a
more extended discussion can be found in more detailed reviews (Bhattacharjee
and Sigl 2000; Berezinsky, Blasi and Vilenkin 1998).

\subsection{Ordinary strings}

Strings can generate UHECRs if there are configurations in which microscopic 
or macroscopic portions of strings annihilate. In the contact regions
the phase of the field trapped in the string becomes undetermined and the
vacuum expectation value becomes non zero. If $\eta\sim m_X$ is the symmetry 
breaking scale at which the string formed, it is easy to see that during
intercommutation of strings or self-intersection, only one (or a few) 
X-particles are generated. It was shown (Shellard 1987; Gill and Kibble 1994) 
that self-intersection events provide a flux of UHECRs which is much smaller 
than required. The same conclusion holds for intercommutation between strings. 

The efficiency of the process can be enhanced by multiple loop fragmentation: 
as a nonintersecting closed loop oscillates and radiates its energy away, the
loop configuration gradually changes. After the loop has lost a substantial 
part of its energy, it becomes likely to self-intersect and fragment into
smaller and smaller loops, until the typical size of a loop 
becomes comparable with the string width $\eta$. At this point the energy 
is radiated in the form of X-particles. Although the process of loop 
fragmentation is not well known, some analytical approximations (Berezinsky,
Blasi and Vilenkin 1998) show that appreciable UHECR fluxes imply
utterly large gamma ray cascade fluxes. Battacharjee and Sigl (2000) 
argued however that there might be models of loop fragmentation in which 
this result is mitigated.

Another way of liberating X-particles is through cusp annihilation 
(Brandenberger 1987). Cusps can be produced along a string loop 
(Turok 1984) or due to kinks propagating in opposite directions 
on a long string (Mohazzab and Brandenberger 1993). Although during the 
cusp annihilation a macroscopic fraction of the string length can be 
transformed into X-particles, the corresponding UHECR flux is far too 
low (Bhattacharjee 1989;Gill and Kibble 1994).

The idea that long strings lose energy mainly through 
formation of closed loops was recently challenged by Vincent, Antunes 
and Hindmarsh (1998). Their simulations seem to show that the string can 
produce X-particles directly and that this process dominates over the
generation of closed loops. This new picture was recently questioned by 
Moore and Shellard (1998).

Even if the results of Vincent et al. (1998) are correct however, they 
cannot solve the problem of UHECRs (Berezinsky, Blasi and Vilenkin 1998): 
in fact the typical 
separation between two segments of a long string is comparable with the 
Hubble scale, so that UHECRs would be completely absorbed.
If by accident a string is close to us (within a few tens Mpc) then 
the UHECR events would appear to come from a filamentary region of 
space, implying a large anisotropy which is not observed. 
Even if the UHECR particles do not reach us the gamma
ray cascade due to absorption of UHE gamma rays produced at large distances 
imposes limits on the efficiency of direct production of X-particles by
strings.

\subsection{Monopolonia and monopole-strings networks}

The role of monopolonia (bound states of monopoles and antimonopoles) 
for UHECRs was first pointed out by Hill (1983) and Schramm and Hill (1983). 
When the monopolonium is produced it is in a very excited state, and later
decays to the ground state, where the typical distance between the monopole
and the antimonopole is smaller than the size of the inner (quantum) stable
orbit. At this point the monopole and the antimonopole annihilate and 
generate X-particles. This process was studied in detail by Bhattacharjee and
Sigl (1995). However, recently Blanco-Pillado and Olum (1999) have found that
plasma friction on the monopoles result in a short lifetime for the
monopolonium, making it useless for UHECRs. 

As an alternative, a similar system was proposed by Berezinsky, Martin and
Vilenkin (1997), where the symmetry breakings 
$G\to H\times U(1) \to H\times Z_N$ ($N\geq 3$) results first in the formation
of monopoles and then of strings connecting them and forming an
infinite network. The shrinking of strings during their evolution causes
the distance between monopoles to decrease. During this stage, monopoles
accelerate and therefore radiate very high energy gluons, that generate
hadrons through fragmentation. The fluxes of UHECRs that result from this 
process are negligible (Berezinsky, Blasi and Vilenkin 1998).

\subsection{Necklaces}

An interesting special case of monopole-string networks is realized when 
the following chain of symmetry breakings occurs: 
$G\to H\times U(1) \to H\times Z_2$. In this case each monopole gets attached
to two strings, to form a necklace (Berezinsky and Vilenkin 1997).
\begin{figure}[thb]
 \begin{center}
  \mbox{\epsfig{file=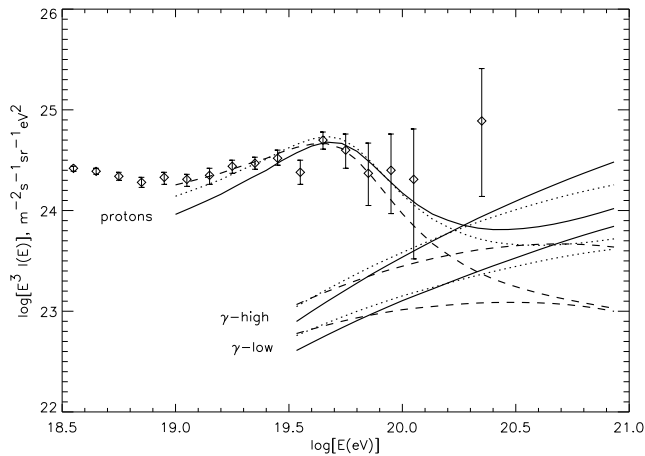,width=9.cm}}
  \caption{\em {Spectrum of UHECRs produced by necklaces. 
}}
 \end{center}
\end{figure}
The critical parameter that defines the dynamics of this network is the 
ratio $r=m/\mu d$ where $m$ is the monopole mass and $d$ is the typical 
separation between monopoles (e.g. the length of a string segment).
Berezinsky and Vilenkin (1997) proposed that there might be cases in
which the system evolves toward a state where $r\gg 1$, although only
numerical simulations can confirm this point. 
In this case, the distance between the monopoles decreases and in the 
end the monopoles annihilate, with the production of X-particles and
their decay to UHECRs. The rate of generation of X-particles is easily found
to be ${\dot n}_X\sim r^2\mu/t^3 m_X$. The quantity $r^2\mu$ is upper 
limited by the cascade radiation, given by $\omega_{cas}=\frac{1}{2} 
f_\pi r^2 \mu =\frac{3}{4} f_\pi r^2\mu/t_0^2$ ($f_\pi\sim 0.5-1$).
The typical distance from the Earth at which the monopole-antimonopole 
annihilations occur is comparable with the typical separation between 
necklaces, $D\sim\left(\frac{3f_\pi\mu}{4t_0^2\omega_{cas}}\right)^{1/4} > 10 
(\mu/10^6 GeV^2)^{1/4}$ kpc.

Clearly, necklaces provide an example in which the typical separation between 
defects is smaller than the pathlength of gamma rays and protons at 
ultra-high energies.
Hence necklaces behave like a homogeneous distribution of sources, so that
the proton component has the usual GZK cutoff. This 
component dominates the UHECR flux up to $\sim 10^{20}$ eV, while 
at higher energies gamma rays take over. The fluxes obtained 
in (Berezinsky, Blasi and Vilenkin 1998) are reported in Fig. 2, where the
SUSY-QCD fragmentation functions (Berezinsky and Kachelriess 1998) were used. 
The dashed lines are for $m_X=10^{14}$ GeV, the dotted lines for $m_X=
10^{15}$ GeV and the solid lines for $m_X=10^{16}$ GeV.
The two curves for gamma rays refer to two different assumptions about the 
radio background at low frequencies (Protheroe and Biermann 1996).

\section{Cosmological relic particles}

Super heavy particles with very long lifetime can be produced in the early 
universe and generate UHECRs at present. The existence of
particles satisfying these requirements was studied recently by Berezinsky,
Kachelriess and Vilenkin (1997); Kuzmin and Rubakov (1997), Chung,
Kolb and Riotto (1998) and Kuzmin and Tkachev (1998) and reviewed 
by Kuzmin and Tkachev (1998a). In order to keep the same symbolism used 
in previous sections, we will call these particles X-particles.

X-particles can be produced in the early universe through different
mechanisms. The simplest of them is the {\it gravitational production}:
particles are produced naturally in a time variable gravitational field
or indeed in a generic time variable classical field. In the gravitational
case no additional coupling is required (all particles interact 
gravitationally). If the time variable field is the inflaton field $\phi$, 
a direct coupling of the X-particles to $\phi$ is needed. 

The gravitational production of particles was first proposed by Zeldovich
and Starobinsky (1972). It does not require any additional assumption
neither on the X-particles nor on cosmology. In particular inflation is 
not required a priori, and indeed it reduces the effect. It can be shown 
that at time t, gravitational production can only generate X-particles 
with mass $m_X\leq H(t)\leq  m_\phi$, where $H(t)$ is the Hubble constant
and $m_\phi$ is the inflaton mass. Chung, Kolb and Riotto (1998) and 
Kuzmin and Tkachev (1998) demostrated the impressive result that the fraction 
of the critical mass contributed by X-particles with $m_X\sim 10^{13}$ GeV 
produced gravitationally is $\Omega_X\sim 1$, with no additional assumption! 
In other words, cold dark matter can naturally be explained in terms of 
X-particles in this range of masses.

If the X-particles are directly coupled to the inflaton field, they can be 
effectively generated during preheating (Kofman, Linde and Starobinsky 1994;
Felder, Kofman and Linde 1998). Alternative mechanisms for the production of 
X-particles are based on non-equilibrium thermal generation during the 
preheating stage (Berezinsky, Kachelriess and Vilenkin 1997).

As mentioned in the beginning of this section, in order for X-particles 
to be useful dark matter candidates and generate UHECRs they need to be long 
lived. The gravitational coupling by itself induces a lifetime much shorter 
than the age of the universe for the range of masses which we are interested 
in. Therefore, in order to have long lifetimes, additional symmetries must be 
postulated: for instance discrete gauge symmetries can protect X-particles 
from decay, while being very weakly broken, perhaps by instanton effects 
(Kuzmin and Rubakov 1998). These effects can allow decay times larger than 
the age of the universe, as shown by Hamaguchi, Nomura and Yanagida (1998).

The slow decay of X-particles produces UHECRs.
The interesting feature of this model is 
that X-particles cluster in the galactic halo, as cold dark matter 
(Berezinsky, Blasi and Vilenkin 1998). [If monopolonia survived they 
would also cluster in the halo]. 
Hence UHECRs are expected to be produced locally, with no absorption, and as 
a consequence the observed spectra are nearly identical to the emission 
spectra, and therefore gamma rays dominate. The very flat 
spectra and the gamma ray composition are two of the signatures.
The calculations of the expected fluxes have been performed by Berezinsky, 
Blasi and Vilenkin (1998), Birkel and Sarkar (1998) and Blasi (1999). In 
figure 3 we report the results found in (Blasi 1999). The two solid lines 
are obtained for SUSY-QCD fragmentation functions and the dashed lines are 
for the QCD fragmentation function in the MLLA approximation, with 
$m_X=10^{14}$ GeV (solid lines) and $m_X=10^{13}$ GeV (thin lines). 
\begin{figure}[thb]
  \begin{center}
  \mbox{\epsfig{file=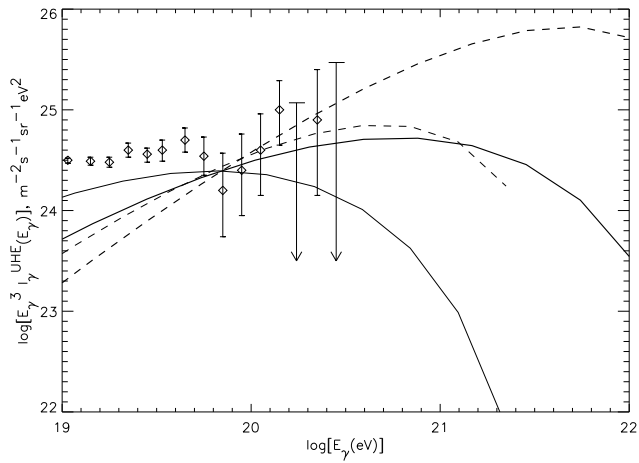,width=9.cm}}
  \caption{\em {Spectrum of UHECRs produced by SH particles in the halo. 
}}
 \end{center}
\end{figure}

The strongest signature of the model is a slight anisotropy due to the 
asymmetric position of the sun in the Galaxy (Dubovsky and Tinyakov 1998;
Berezinsky, Blasi and Vilenkin 1998). 
More recently a detailed evaluation of the amplitude and phase of the first 
harmonic has been carried out by Berezinsky and Mikhailov (1999) and 
Medina Tanco and Watson (1999). The two papers agree that the present
data is consistent with the anisotropy expected in the model of X-particles
in the halo. 
In fact, as discussed in section 2, observations at 
present do not suggest any appreciable deviation from isotropy, with the 
exception of a few degree scale anisotropies showing up in the form of 
doublets and triplets of events within an angular scale comparable to the 
resolution of the experiments. Uchihori et al. (2000) investigated the total 
of 92 events above $4\times 10^{19}$ eV collected by Volcano Ranch, Haverah 
Park, Yakutsk and Akeno and found 12 doublets and 3 triplets within $3^o$.

In TD models the presence of these multiplets of events is usually not well 
accomodated because of the homogeneous distribution of the topological
defects or of the superheavy particles. However, it was shown by Blasi and
Sheth (2000) that in the latter the presence of the multiplets is actually 
naturally explained, taking into account the dark matter distribution in the 
galactic halo. 

The spatial distribution of cold dark matter has been studied in detail in 
N-body simulations (Ghigna et al. 1999; Moore et al. 1999; Tormen, Diaferio 
and Syer 1998) and a few elements seem to be well established: 1) the 
density of dark matter has a behaviour $r^{-\gamma}$ with $\gamma\sim 1-1.5$ 
in the central part of galaxies; 2) at large radii the radial profile is 
$r^{-3}$; 3) in addition to the smooth component, simulations show very 
clearly the existence of a clumped component. 
Berezinsky (2000) suggested that the multiplets might be correlated with 
the clumps of dark matter.
Blasi and Sheth (2000) carried out a numerical simulation of the arrival 
directions of UHECR events from a realistic distribution of dark matter in 
the halo (including the clumps) and derived the probabilities to detect the 
observed numbers of doublets and triplets (or more). 
They adopted a Navarro-Frenk-White (NFW: Navarro, Frenk and White 1996) 
smooth distribution of 
dark matter and used the results of the N-body simulations for the small 
scale structure, assuming each clump being represented by an isothermal 
sphere. 
The numbers of doublets and triplets found by generating a large number of 
halo configurations and the corresponding pattern of arrival directions of 
UHECRs are comparable with the observed ones and in excess of the number
expected from an isotropic distribution of arrival directions.
Surprisingly, the main reason for the increase in the predicted number of 
doublets and triplets with respect to isotropy is that the 
smooth component has an NFW profile. The effect of the clumps, apart from 
increasing slightly the numbers of multiplets, is to enhance the probability 
of having more than the average number of doublets and triplets. 
This result can be understood by accounting for the shape of the dark matter 
profile, increasing toward the central part of the galaxy. Predictions for 
future full sky experiments were proposed.

\section{Conclusions}

The problem of the existence of UHECRs is all but solved. Astrophysical
models are being narrowed down by several constraints: most of the 
extragalactic sources would suffer a GZK cutoff, since the large 
overdensities needed locally to avoid it do not seem to be present.
Moreover it is difficult to reach the highest energies (Norman, Melrose and
Achterberg 1995). 
Galactic accelerators usually require an iron dominated composition
in order to account for the isotropy of arrival directions. Further 
studies are needed to quantify the expected and observed anisotropies in 
these models, since the magnetic field strength and extension in the halo
are poorly known. 

Top-Down models naturally provide the highest energies, and at least some
of them can describe quite well the observed sprectal shape above 
$\sim 4\times 10^{19}$ eV. As positive examples we considered in some
detail the cases of necklaces and relic super massive particles in 
the halo of the Galaxy. In the former model, the composition of UHECRs
should be dominated by protons up to $\sim 10^{20}$ eV, and by photons
at higher energies. In the latter, UHECRs are gamma rays and a peculiar
pattern of large and small scale anisotropies should be visible in future
observations. In all Top-Down models there is at least an appreciable fraction
of UHECRs in the form of gamma rays.

Experiments like HiRes (Corbato' et al. 1992), currently operating, the 
Auger project (Cronin 1992), the proposed Telescope Array (Teshima et al. 
1992) and the OWL-Airwatch satellite (Streitmatter 1997) will nail down 
the answers to the dark points. An unambiguous determination of the composition
will be fundamental: heavy composition would rule out all Top-Down models
and open a window of opportunity for galactic scenarios. A Gamma ray dominated
composition would instead be a smoking gun for Top-Down models. 
The measurement of the anisotropy on different scales will also be a crucial
step: galactic models all predict some degree of anisotropy toward the galactic
disk or center. A peculiar pattern of anisotropy is also predicted by super
heavy particles clustered as dark matter in the galactic halo. 
 
{\bf Aknowledgments} This work was supported by the DOE and the 
NASA grant NAG 5-7092 at Fermilab.

\section { References}

\reff Ahn, E.J., Biermann, P.L., Medina-Tanco, G., Stanev, T.: 1999, preprint
astro-ph/9911123.

\reff
Berezinsky, V.S.: 1999 Invited talk at TAUP-99, Paris, September 6 - 10, 1999, 
preprint hep-ph/0001163.

\reff Berezinsky, V.S., Blasi, P., Vilenkin, A.: 1998 Phys. Rev. {\bf D58},
p. 103515.

\reff Berezinksy, V.S., Bulanov, S.V., Dogiel, V.A., Ginzburg, V.L.,
Ptuskin, V.S.: 1990 {\it Astrophysics of Cosmic Rays} (Amsterdam: North
Holland, 1990).

\reff Berezinsky, V.S., Kachelriess, M.: 1998 Phys. Lett. {\bf B434}, p. 61.

\reff Berezinsky, V.S., Kachelriess, M., Vilenkin, A.: 1997 Phys. Rev. Lett. 
{\bf 79}, p. 4302.

\reff Berezinsky, V.S., Martin, X., Vilenkin, A.: 1997 Phys. Rev. {\bf D56},
p. 2024.

\reff 
Berezinsky, V.S., Mikhailov, A.: 1999 Phys. Lett. {\bf B449},p.  237.

\reff Berezinsky, V.S., Vilenkin, A.: 1997 Phys. Rev. Lett. {\bf 79}, p. 5202.

\reff Bhattacharjee, P.: 1989 Phys. Rev. {\bf D40}, p. 3968.

\reff Bhattacharjee, P., Rana, N.C.: 1990 Phys. Lett. {\bf B246}, p. 365.

\reff Bhattacharjee, P., Sigl, G.: 2000 Phys. Rep. {\bf 327}, p. 109.

\reff Bhattacharjee, P., Sigl, G.: 1995 Phys. Rev. {\bf D51}, p. 4079.

\reff Bird, D.J., et al.: 1995 Astrophys. J. {\bf 441}, p. 144; 1993 Phys. 
Rev. Lett. {\bf 71}, p. 3401; 1994 Astrophys. J. {\bf 424}, p. 491.

\reff 
Birkel, M., Sarkar, S.: 1998 Astropart. Phys. {\bf 9}, p 297. 

\reff Blanco-Pillado, J.J., Olum, K.D.: 1999, preprint astro-ph/9904315.

\reff Blanton, M., Blasi, P., Olinto, A.V.: 2000, in preparation.

\reff Blasi, P., Burles, S., Olinto, A.V.: 1999 Astrophys. J. Lett. 
{\bf 512}, p. L79.

\reff Blasi, P., Olinto, A.V.: 1999 Phys. Rev. {\bf D59}, p. 023001.

\reff Blasi, P., Sheth, R.K.: 2000 preprint astro-ph/0006316, Phys. Lett.
B, in press.

\reff Blasi, P.: 1999 Phys. Rev. {\bf D60}, p. 023514.

\reff Brandenberger, R.: 1987 Nucl. Phys. {\bf B293}, p. 812.

\reff Chung, D.J.H., Kolb, E.W., Riotto, A.: 1998 Phys. Rev. {\bf D59}, 
p. 023501.

\reff Corbato',et al.: 1992 Nucl. Phys. B. (Proc. Suppl.) {\bf 28B}, p. 36.

\reff Cronin, J.W.: 1992 Nucl. Phys. B. (Proc. Suppl.) {\bf 28B}, p. 213.

\reff Dubovsky, S.L., Tynyakov, P.G.: 1998 Pis'ma Zh. Eksp. Teor. Fiz. 
{\bf 68},p. 99 [JETP Lett. {\bf 68}, p. 107].

\reff Efimov, N.N., et al.: 1991 Ref. Proc. International Symposium on 
{\it Astrophysical Aspects of the most energetic cosmic rays}, eds M. Nagano
and F. Takahara (World Scientific, Singapore), p. 20.

\reff Elbert, J.W., Sommers, P.: 1995 Astrophys. J. {\bf 441}, p. 151.

\reff Felder, G., Kofman, L., Linde, A.: 1998 preprint astro-ph/9812289.

\reff Ghigna, S., Moore, B., Governato, F., Lake, G., Quinn, T., Stadel, J.:
1999 preprint astro-ph/9910166.

\reff Gill, A.J., Kibble, T.W.B.: 1994 Phys. Rev. {\bf D50}, p. 3660.

\reff Greisen, K.: 1966 Phys. Rev. Lett. {\bf 16}, p. 748.

\reff Halzen, F., Vazques, R., Stanev, T., Vankov, H.S.: 1995 Astropart. 
Phys. {\bf 3}, p. 151

\reff Hamaguchi, K., Nomura, Y., Yanagida, T.: 1998 Phys. Rev. {\bf D58},
p. 103503. 

\reff Hayashida, N., et al.: 1994 Phys. Rev. Lett. {\bf 73}, p. 3491.

\reff Hill, C.T.: 1983 Nucl. Phys. {\bf B224}, p. 469.

\reff Hill, C.T., Schramm, D.N., Walker, T.P.: 1987 Phys. Rev. {\bf D36}, p. 
1007.

\reff Kieda, D., et al. {\it HiRes Collaboration}: 1999 Proc. of 26th ICRC,
Salt Lake City, Utah.

\reff Kofman, L., Linde, A., Starobinsky, A.: 1994 Phys. Rev. Lett. {\bf 73},
p. 3195.

\reff Kronberg, P.P.: 1994 Rep. Prog. Phys. {\bf 57}, p. 325.

\reff Kuzmin, V.A., Rubakov, V.A.: 1998 Yadern. Fiz. {\bf 61}, p. 1122.

\reff  Kuzmin, V.A., Tkachev, I.I.: 1998 JETP Lett. {\bf 69}, p. 271.

\reff Kuzmin, V.A., Tkachev, I.I.: 1998a preprint hep-ph/9903542.

\reff Lawrence, M.A., Reid, R.J.O., Watson, A.A.: 1991 J. Phys. G. Nucl.
Part. Phys. {\bf 17}, p. 773.

\reff Lee, S., Olinto, A.V., Sigl, G.: 1995 Astrophys. J. Lett. {\bf 455}, 
p. L21.

\reff Lemoine, M., Sigl, G., Olinto, A.V., Schramm, D.N.: 1997 Astrophys. J. 
Lett. {\bf 486}, p. L115.

\reff Linsley, J.: 1963 Phys. Rev. Lett. {\bf 10}, p. 146.

\reff Masperi, L., Silva, B.: 1998 Astropart. Phys. {\bf 8}, p. 173.

\reff 
Medina-Tanco, G.A., Watson, A.A.: 1999 Astropart. Phys. {\bf 12},p. 25.

\reff Mohazzab, M., Brandenberger, R.: 1993 Int. Journ. Mod. Phys. {\bf D2},
p. 183.

\reff Moore, B., Ghigna, S., Governato, F., 
Lake, G., Quinn, T., Stadel, J., Tozzi, P.: 1999 Astrophys. J. {\bf 524},p. 19.

\reff Moore, J.N., Shellard, E.P.S.: preprint astro-ph/9808336.

\reff Navarro, J.F., Frenk, C.S., White, S.D.M.: 1996 Astrophys. J. 
{\bf 462}, p. 563.

\reff Norman, C.T., Melrose, D.B., Achterberg, A.: 1995 Astrophys. J. 
{\bf 454}, p. 60.

\reff Olinto, A.V.: 2000 preprint astro-ph/0002006, to be published in the 
David Schramm Memorial Volume of Physics Reports.

\reff Protheroe, R.J., Biermann, P.L.: 1996 Astropart. Phys. {\bf 6}, p. 45.

\reff Schramm, D.N., Hill, C.T.: 1983 Proc. 18th ICRC (Bangalore) 
{\bf 2}, p. 393.

\reff Shellard, E.P.S.: 1987 Nucl. Phys. {\bf B283}, p. 624.

\reff Sigl, G., Lemoine, M., Biermann, P.L.: Astropart. Phys. {\bf 10}, 
p. 141.

\reff Streitmatter, R.E.: 1997 Proc. of {\it Workshop on Observing Giant 
Cosmic Air Showers from $>10^{20}$ eV Particles from Space}, eds. 
Krizmanic, J.F., Ormes, J.F., and Streitmatter, R.E. (AIP Conference
Proceedings 433, 1997).

\reff Takeda, M., et al.: 1998 Phys. Rev. Lett. {\bf 81}, p. 1163.

\reff Takeda, M. et al.: 1999 Astrophys. J. {\bf 522}, p. 225.

\reff Teshima, M., et al.: 1992 Nucl. Phys. B. (Proc. Suppl.) {\bf 28B}, p. 
169.

\reff
Tormen, G., Diaferio, A., Syer, D.: 1998 MNRAS, {\bf 299},p.  728.

\reff Turok, N.: 1984 Nucl. Phys. {\bf B242}, p. 520.

\reff Uchihori, Y., Nagano, M., Takeda, M., Teshima, M., Lloyd-Evans, J., 
Watson, A.A.: 2000 Astropart. Phys. {\bf 13}, p. 151.

\reff Vilenkin, A., Shellard, E.P.S.: 1994 {\it Cosmic Strings and Other
Topological Defects}, Cambridge University Press, Cambridge.

\reff Vincent, G., Antunes, N., Hindmarsh, M. : 1998 Phys. Rev. Lett. 
{\bf 80}, p. 2277.

\reff Yoshida, S., Dai, H.: 1998 J. Phys. G {\bf 24}, p. 905.

\reff Zatsepin, G.T., Kuzmin, V.A.: 1966 Sov. Phys. JETP Lett. {\bf 4}, 
p. 78.

\reff Zeldovich, Ya.B., Starobinsky, A.A.: 1972 Soviet Phys. JETP {\bf 34},
p. 1159. 

\end{document}